\documentclass[12pt,preprint,aps,prb,showpacs,floatfix]{revtex4}

\usepackage{graphicx}
\usepackage{latexsym}
\usepackage{dcolumn}

\begin{document}

\title{Calculating and visualizing the density of states \\
for simple quantum mechanical systems}

\author{Declan \surname{Mulhall}}
\email{mulhalld2@scranton.edu}
\affiliation{Department of Physics/Engineering\\
University of Scranton\\
Scranton, Pennsylvania 18510-4642, USA}

\author{Matthew J. \surname{Moelter}}
\email{mmoelter@calpoly.edu}
\affiliation{Department of Physics \\
California Polytechnic State University\\
San Luis Obispo, California 93407, USA}

\begin{abstract}
We present a graphical approach to understanding the degeneracy,
density of states, and cumulative state number for some simple
quantum systems. By taking advantage of basic computing operations
we define a straightforward procedure for determining the
relationship between discrete quantum energy levels and the
corresponding density of states and cumulative level number. The
density of states for a particle in a rigid box of various shapes
and dimensions is examined and graphed. It is seen that the
dimension of the box, rather than its shape, is the most important
feature. In addition, we look at the density of states for a
multi-particle system of identical bosons built on the single-particle spectra of those boxes. A simple model is used to explain
how the $N$-particle density of states arises from the single
particle system it is based on.
\end{abstract}

\pacs{24.60.-k,24.60.Lz,25.70.Ef,28.20.Fc}

\maketitle

\section{\label{sec:intro}Introduction}

The concept of the density of states (DOS) is used in many areas of
physics.  For example, it is important for reaction rates in nuclear
physics, the calculation of specific heat capacities, black-body
radiation, phonon spectra, and so on.
\cite{Schroeder,Baierlein,Riedi,Reif,Mandl,Kittel} The DOS arises
naturally and early  in statistical physics. To calculate average
quantities in statistical physics, one could do an integral over
phase space, but this is typically very complex. The alternative is
to express the variable of interest in terms of the energy of the
system. The volume element in phase space is replaced by a weighting
factor in an energy integral, which is often much easier to work
with. This weighting factor is the density of states
 and is the subject of this paper.

Consider the problem of calculating $\langle f\rangle$, the expected
outcome of a measurement of some physical quantity $f$. If you know
the allowed quantum states of the system and can calculate $f_i$,
the value of $f$ in the $i$th state, then $\langle f\rangle = \sum_i
f_i \,P_i$, where $P_i$ is the probability of the system being in
the $i$th state. This is sometimes referred to as a sum over
microstates. It is often easier to write $f$ as a function of energy
and perform a sum over the allowed energies. We then have $\langle
f\rangle = \sum_i f(\epsilon_i) \,P(\epsilon_i)$, where the
probability of the system being on the energy level $\epsilon$ is
given by $P(\epsilon)=d_\epsilon\,e ^{-\beta \epsilon}/{Z}$. Here
$Z=\sum_j e ^{-\beta \epsilon_j}$ is the partition function, $\beta
= 1/kT$, and $d_\epsilon$ is the degeneracy of the level or number of
states with energy $\epsilon$.  If the system has an energy
$\epsilon$, then each of these $d_\epsilon$ states is equally
likely. Now, for our expected value of $f$ we have
\begin{equation}
\sum_{i} f_i\,P_i\rightarrow \sum_{\epsilon}d_\epsilon \,
f(\epsilon)\,P(\epsilon).
\end{equation}

In a system where $\epsilon$ is continuous (or effectively so), the
sum becomes an integral and $d_\epsilon$ is replaced by
$g(\epsilon)$---the DOS---a measure of how many states there are in a
small range $d\epsilon$ around $\epsilon$. The DOS $g(\epsilon)$ is no longer an
absolute number of states; it is now a weighting factor
in an integral over energy. We must be careful here because
$g(\epsilon)$ is often confused with the level density; indeed, the
terms are often used interchangeably. We use the term ``level" to
mean an allowed value of energy  and the DOS is the level density
multiplied by a degeneracy factor.

In this paper we present a procedural approach to obtain the DOS for
systems that arise in modern physics and statistical mechanics
courses. In the next section we apply the procedure to a single
particle in a rigid box.  In Sec.~\ref{sec:procedure} the
procedure for visualizing the DOS is summarized and the effect of
the shape of the box on the DOS is examined. Section~\ref{sec:npart}
addresses systems of $N$ noninteracting bosons, where $N$-particle spectra
are calculated and the dependance of their DOS on $N$ is compared
with a simple model.

\section{\label{sec:box} Visualizing the density of states for a particle in a box}

The particle in a box is one of the first examples  students
encounter in quantum mechanics. It is simple enough to solve from
scratch by hand and exhibits much of the salient nonclassical
behavior. Furthermore, it serves as a basic template for a host of
interesting topics: scattering, double-well potentials, and
perturbation theory  to name a few. We will discuss the DOS in the
context of a particle in a box of various dimensions. First we
introduce the concept of degeneracy via numerical results for the
spectrum. This leads naturally to the idea of a cumulative state
number from which the DOS naturally follows. Then an analytic
approach to obtaining the DOS is presented.

We start with a (nonrelativistic) particle of mass $M$ inside a box.
We assume the box is rigid, by which we mean that the potential
energy is infinite outside the box and zero inside the box. In a
one-dimensional box of length $L$ the energy levels are given by
\cite{Serway} $\epsilon_n=\epsilon_0n^2$, where
$\epsilon_0=\pi^2\hbar^2/(2ML^2)$ and the quantum number $n$ is an
integer. In a two-dimensional square box with sides of length $L$,
the energy levels are given by $\epsilon_n=\epsilon_0(n_x^2+n_y^2)$,
where  $n_x$ and $n_y$ are the quantum numbers corresponding to the
two spatial dimensions. In the three-dimensional case of a cubical
box of side length $L$, we have $\epsilon_n=\epsilon_0(n_x^2+n_y^2+n_z^2)$.

\begin{figure}[h!]
\includegraphics[width=8.5cm]{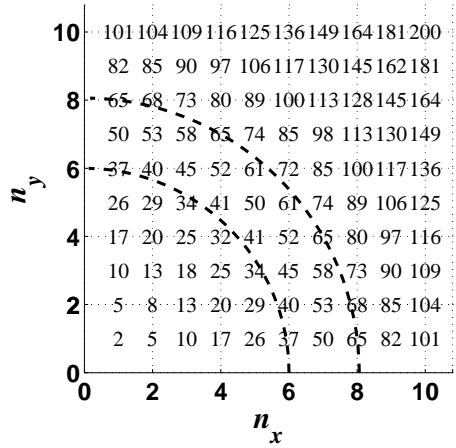}
\caption{\label{fig:2DCircle} The energies for different
combinations of $n_x$ and $n_y$ for a particle in a 2-D square box. The numbers at the
grid sites are $n_x^2 + n_y^2$. The quarter circles represent constant
$n=\sqrt{n_x^2 + n_y^2}$; here we show $n=\sqrt{36}$ and $n=\sqrt{65}$.
Certain values of $n$ give circles that intersect with the grid
points. These correspond to the allowed energies $\epsilon = n^2
\epsilon_0$. Notice that $n=6$ does not intersect any grid point, while
$n=\sqrt{65}\approx 8$ (dashed line) intersects four grid points,
corresponding to the degeneracy of that energy level.}
\end{figure}

It is natural to associate each quantum number with a number line
and each integer value with a point along this line. In two
dimensions the pair of quantum numbers defines a plane or
two-dimensional space called $n$-space.\cite{CircleRefs} This space
could also be called momentum space by using the identity $p_x=\hbar
k_x = n_x \pi \hbar / L$. The energy depends on the sum of the
squares of $n_x$ and $n_y$.  One way of thinking about the
distribution of these energies is to locate them in $n$-space. Using
the horizontal axis for $n_x$ and the vertical axis for $n_y$, we
can write all energy values at their corresponding grid points. This
is done in Fig.~\ref{fig:2DCircle}, where at each pair of quantum
numbers $(n_x,n_y)$ the energy is written (in units of
$\epsilon_0$); for example, at grid points $(n_x,n_y)=(1,3)$ and
$(3,1)$ we see a ``10." Suppose $n$ is the radius of a circle in
$n$-space such that $n^2=n_x^2+n_y^2$. Rewriting the energy as
$\epsilon_n=\epsilon_0 n^2$ we see that in $n$-space a given energy
(in units of $\epsilon_0$) corresponds to a circle of radius
$\sqrt{\epsilon/\epsilon_{0}}$. (The radius $n$ is related to the
magnitude of the momentum vector by $|\vec{p}|=n\pi\hbar/L$). In
Fig.~\ref{fig:2DCircle} the circles corresponding to energies ``36'' and ``65"
are shown as dotted curves with radii $\sqrt{36} = 6$ and $\sqrt{65} \approx 8$, respectively.

When energies correspond to more than one independent state we say they are ``degenerate.''
If we make a list of all energies corresponding to the various
quantum numbers $(n_x,n_y)$ and order them by energy, we can make a
plot of the ``number of states with energy $\epsilon$'' vs
$\epsilon$. This quantity is $d_\epsilon$, the degeneracy of the
energy $\epsilon$, and for discrete spectra it is an integer. A plot
of $d_\epsilon$ is a series of spikes, of height $d_\epsilon$, at
each allowed $\epsilon$.

\begin{figure}[h!]
\includegraphics[width=8.5cm]{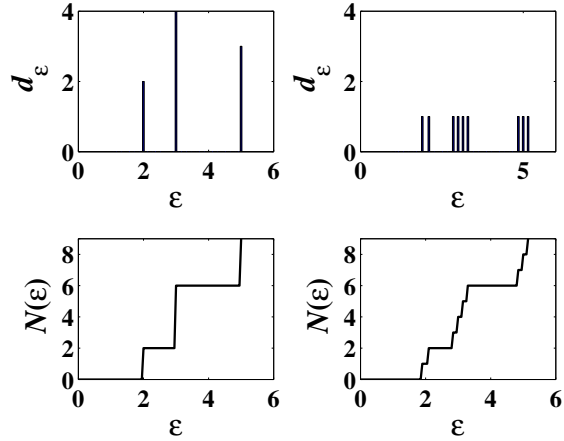}
\caption{\label{fig:Degeneracy}The degeneracy $d_{\epsilon}$ (upper
panels) and corresponding cumulative state number ${\mathcal
N}(\epsilon)$ (lower panels) for the lowest nine levels of a
simple, discrete system. On the left, the degeneracy
 of the system is intact while on the right, the degeneracy has
been lifted. The effect on $d_{\epsilon}$ is dramatic; less
so for ${\mathcal N}(\epsilon)$.}
\end{figure}

To further illustrate this idea, imagine a hypothetical single-particle spectrum
where the lowest nine energies are
 $\{\epsilon\}=\{2,2,3,3,3,3,5,5,5\}$. We have plotted $d_\epsilon$
vs\ $\epsilon$ for this system in Fig.~\ref{fig:Degeneracy} (upper left panel). At this
point it is helpful to introduce the cumulative state number ${\mathcal N}(\epsilon)$, defined as the number of states
with energy $\le \epsilon$; its graph is a staircase where each
step has a height $d_\epsilon$ and a width determined by the gap to
the next energy (lower left panel of Fig.~\ref{fig:Degeneracy}). In other words,
given an ordered list of energies $\{\epsilon_i\}$ we have
${\mathcal N}(\epsilon)= i \quad\mathrm{ for }\quad \epsilon_i <
\epsilon < \epsilon_{i+1}$.


A plot of the spikes  $d_\epsilon$ gives a visual measure
of the degeneracies of the energies. These degeneracies are
quite delicate in the sense that most perturbations
to the potential will break them and the picture for
$d_\epsilon$ will change dramatically. Each $d_\epsilon$-high
spike will turn into a cluster of $d_\epsilon$ separate
spikes, each one unit high. The spacing of the spikes will be
determined by the strength of the perturbation. The corresponding
change in ${\mathcal N}(\epsilon)$ is that each step in the
unperturbed system that was $d_\epsilon$ high will now
become a series of short steps, each one unit high
(upper and lower right panels of Fig.~\ref{fig:Degeneracy}).
It is hoped that Figs.~\ref{fig:2DCircle}
and~\ref{fig:Degeneracy} will be a useful starting point for
student discussion.

A smooth DOS function $g(\epsilon)$ is useful because at higher
energies one is interested in the number of states in an interval, or
the density of spikes along the $\epsilon$-axis. The DOS and
cumulative state number are related by
$g(\epsilon)d\epsilon=d\mathcal{N}(\epsilon)$, so the DOS is
recognized as the slope of ${\mathcal N}(\epsilon)$. We must be
careful here because the slope of a staircase is infinite at each
step, so we mean slope in an averaging sense.

To get $g(\epsilon)$ from $\mathcal N(\epsilon)$
we can take a numerical derivative using a
finite difference scheme,
\begin{equation}
\label{eqn:deriv} g(\epsilon)=\frac{d{\mathcal
N}(\epsilon)}{d\epsilon}= \frac{{\mathcal
N}(\epsilon+\delta\epsilon/2)-{\mathcal
N}(\epsilon-\delta\epsilon/2)} {\delta\epsilon}.
\end{equation}
It is worth saying explicitly that if the energies are put
into bins of width $\delta\epsilon$, with center $\epsilon$, then
$g(\epsilon)$=(number of states in
bin)/(width of bin).

\begin{figure}[h!]
\includegraphics[width=8.5cm]{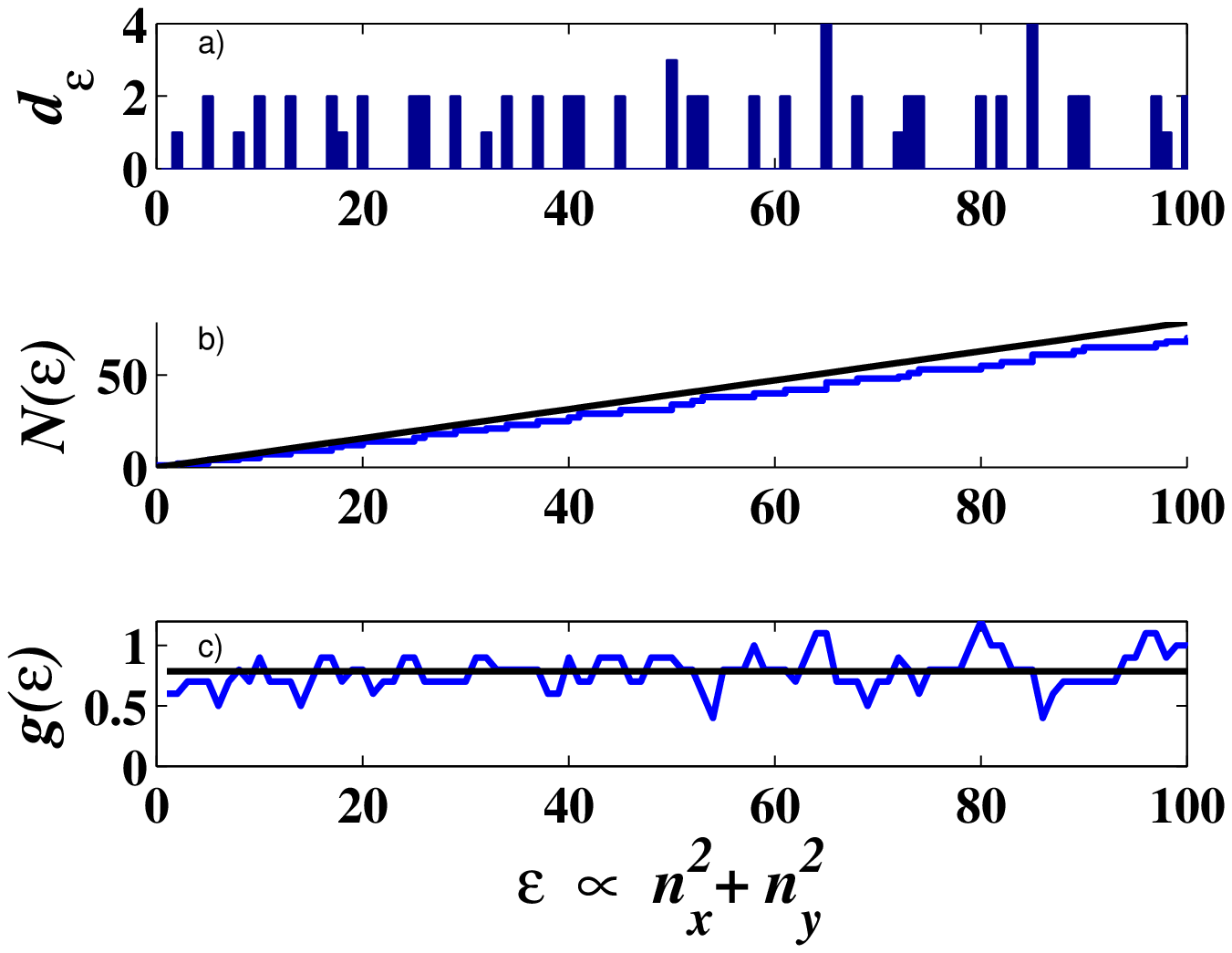}
\caption{\label{fig:2DHistosmall} Top: Degeneracy $d_\epsilon$ of the
allowed energies $\epsilon$ (in units of $\epsilon_{0}$) of a single
particle in a 2-D square box, up to $n_x=n_y=10$. Middle: The
corresponding ${\mathcal N}(\epsilon)$. Theory (smooth line) gives
${\mathcal N}(\epsilon)=(\pi/4)\epsilon$. Bottom: A plot of
$g(\epsilon)$ obtained using the derivative scheme in
Eq.~(\ref{eqn:deriv}) with a window width of 10. Theory (smooth
line) gives $g(\epsilon)=\pi/4$.}
\end{figure}

\begin{figure}[h!]
\includegraphics[width=8.5cm]{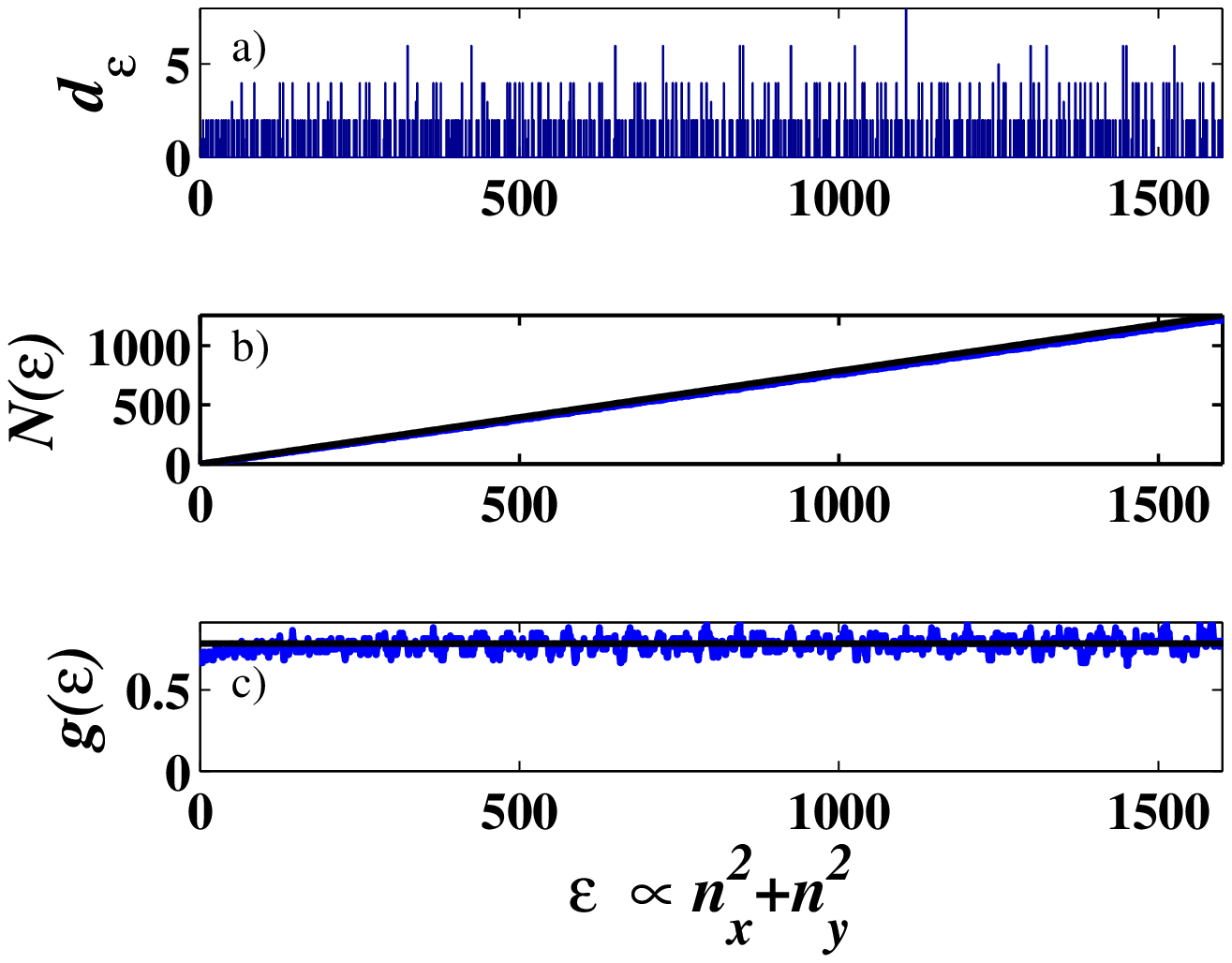}
\caption{\label{fig:2DHistobig} Top: Degeneracy $d_\epsilon$ of the
allowed energies $\epsilon$ (in units of $\epsilon_{0}$) of a single
particle in a 2-D square box, up to $n_x=n_y=40$. Middle: The
corresponding ${\mathcal N}(\epsilon)$. Theory (smooth line) gives
${\mathcal N}(\epsilon)=(\pi/4)\epsilon$. Bottom: A plot of
$g(\epsilon)$ obtained using the derivative scheme in
Eq.~(\ref{eqn:deriv}) with a window width of 60. Theory (smooth
line) gives $g(\epsilon)=\pi/4$.}
\end{figure}

Now that we have a numerical representation for $g(\epsilon)$, we
would like to get an analytic expression for $g(\epsilon)$ that is valid for
the statistical region where the DOS is large and well approximated
by a smooth function. We will do this for the rigid box potentials
where $\epsilon = n^2$ (in units of $\epsilon_0$). In $n$-space
the states with energy in an interval $d\epsilon$ centered on
$\epsilon$ correspond to a set of points in a spherical shell of
thickness $dn$ with all-positive coordinates. In
Fig.~\ref{fig:2DCircle} the number of states between the quarter
circles of radius $n$ and $n+dn$ is proportional to the area of the
curved band. This result is an approximation, because $n_x$ and $n_y$ are
discrete and there are fluctuations, but we will see
that this is a good approximation
nevertheless. The number of states is the ``volume'' of this shell but
by definition it is also $g(\epsilon)\,d\epsilon$, so in 2-D we
have $g(\epsilon)\,d\epsilon = (1/2)\pi n\, dn$. If $n(\epsilon)$
is a smooth function we can  write
$g(\epsilon)=(1/2)\pi n(\epsilon)\,dn/d\epsilon$. Using
$n(\epsilon)=\sqrt{\epsilon}$ and $dn/d\epsilon = 1/(2
\sqrt{\epsilon})$ we see that the DOS is constant:
$g(\epsilon)=\pi /4$. Since the DOS is the slope of ${\mathcal
N}(\epsilon)$, the function ${\mathcal N}(\epsilon)$ should be well
approximated by the straight line ${\mathcal N}(\epsilon) = (\pi/4)
\epsilon$, as shown in Fig.~\ref{fig:2DHistosmall}. The
numerical results are convincing, particularly where the energy
range increases and the fluctuations about the smooth functions get
smaller (see Fig.~\ref{fig:2DHistobig}).



For a particle in a  3-D rigid box with sides $L_x$, $L_y$, and $L_z$,
we will be working with a 3-D $n$-space and again each point
$(n_x,n_y,n_z)$ corresponds to an allowed value of energy where
$\epsilon=(\pi^2\hbar^2)/(2M)[(n_x/L_x)^2+(n_y/L_y)^2+(n_z/L_z)^2]$.
For a cube of side length $L$ we have $\epsilon=(\pi^2\hbar^2)/(2M L^2)n^2=\epsilon_{0}n^{2}$ with
$n^2=(n_x^2+n_y^2+n_z^2)$. Now the appropriate construction to get
the DOS is a shell of thickness $dn$ and radius $n$ in the
positive octant of a sphere, which leads to $g(\epsilon) = (1/2)\pi n^2
\, dn/d\epsilon$. Again, using $n(\epsilon)=\sqrt{\epsilon}$ and
$dn/d\epsilon = 1/(2 \sqrt{\epsilon})$, we see that the DOS now has
the form  $g(\epsilon)=(\pi /4)\sqrt{\epsilon}$ and ${\mathcal
N}(\epsilon)= (\pi/6)\epsilon^{ 3/2}$.  In
Figs.~\ref{fig:3DHistosmall} and ~\ref{fig:3DHistobig} we compare
the numerical results with these smooth functions for different
energy ranges. Again as the energy increases, the numerical
fluctuations get smaller.

We remark here that if we had $N$ non-interacting independent
particles in a rigid cube, then the total energy would be the sum of
the individual energies, and this sum could be related to the surface [xx do you mean volume? xx]
of a $D=3N$-dimensional hypersphere. This problem is similar to the
case of a single particle in a rigid box in a $D$-dimensional space.
We would need to discuss the volume of a sphere of
radius $R$ in $D$-dimensions, denoted by $V_D$ and given by\cite{Garrod}
\begin{equation}
\label{eqn:vp} V_D=\frac{\pi^{D/2}}{\Gamma(D/2
+1)}\, R^D=C_D\,R^D.
\end{equation}
Here, $\Gamma(D+1)=D!$, giving $C_2=\pi$ and
$C_3=\frac{4}{3}\pi$ as expected. The surface area of this
$D$-sphere is given by ${\mathrm S}_D=D\,C_D\,R^{D-1}$. Where before
we were concerned with the length of a quarter circle of radius $n$
($D=2$) and the area of an octant of a sphere of radius $n$
($D=3$), we now have the area of the positive portion of the
$D$-dimensional sphere, given by $(1/2^D)S_D$ (the need for
$n_x$, $n_y \dots$ to be positive gives a factor of $1/2$
for each dimension). The cumulative state number is given by the
volume of phase space enclosed by the boundary $n=\sqrt{\epsilon}$,
\begin{equation}
\label{eqn:nepd} {\mathcal N}(\epsilon)=\frac{1}{2^D}\,\frac{\pi^{D/2}}{\Gamma(D/2
+1)}\,\epsilon^{D/2},
\end{equation}
and differentiation gives
\begin{equation}
\label{eqn:gepd}
g(\epsilon)=\frac{1}{2^{D+1}}\,D\,\frac{\pi^{D/2}}{\Gamma(D/2
+1)}\,\epsilon^{{(D/2)}-1}.
\end{equation}
The forms of ${\mathcal N}(\epsilon)$ and $g(\epsilon)$ for
some representative $D$ are given in Table \ref{tab:dims}. Later we will see
that the DOS for one particle in $D$ dimensions is proportional to
the DOS of $D$ particles in one dimension, so long as the particles do not
interact.


\section{\label{sec:procedure}
Procedure for calculating and visualizing the density of states}

For any quantum-mechanical system we can determine
the cumulative state number and the corresponding density of
states using the following procedure:
\begin{enumerate}
\item Solve the relevant quantum-mechanical
 problem to get a list of the energies of the
system.  (This can be an analytical expression or a list of energies
obtained numerically.)
\item Sort the  list of  energies to
get a set
$\{\epsilon_1,\,\epsilon_2,\,\epsilon_3,\,\dots,\,\epsilon_n \}$ up
to some maximum energy $\epsilon_\mathrm{max}$.
\item Create the cumulative state number ${\mathcal N}(\epsilon)$
by making the set $\{ \epsilon_1, 1\},\,\{ \epsilon_2,
2\},\,\{ \epsilon_3, 3\},\, \dots\{ \epsilon_n, n\}$.
\item Find the DOS by taking the
numerical derivative of the cumulative state number, as in Eq.~(\ref{eqn:deriv}).
Choose a window size and locate the
window so that its center is at $\epsilon$. The value of
$g(\epsilon)$ is the number of energies that are within that window
divided by the window size. Evaluate this for all locations of the
window, and you now have a list of ordered pairs $\{\epsilon,
g(\epsilon)\}$.
\end{enumerate}

As supplementary materials\cite{EPAPS} we include a \textsc{Matlab}\cite{Matlab} program
used to obtain Figs.~\ref{fig:3DHistosmall} and \ref{fig:3DHistobig}; with appropriate
modifications the program can be used to produce Figs.~\ref{fig:2DHistosmall} and \ref{fig:2DHistobig}.

We will apply this procedure to  calculate and visualize the DOS for
particles in rigid boxes with various geometries.



\subsection{\label{sec:cube}
Cubical box}

We first will consider a cube of side length $L$.

\emph{Step 1}: The energies are $\epsilon=(\pi^2\hbar^2)(n_x^2+n_y^2+n_z^2)/(2M V^{2/3}$), where we
have used $L^{2}=V^{2/3}$ with $V$ the volume of the box. If we set
$(\pi^2\hbar^2)/(2M) = 1$ and $V=1$, then the energies are all
integers.

\emph{Step 2}: Using {\sc Mathematica}\cite{Mathematica} we
determined the lowest 15,954 energy levels. To calculate the
energies we need all the triplets $\{n_x,\,n_y,\,n_z\}$ within the
sphere of radius $n_{\mathrm{max}}$ in $n$-space. We made a list of
$n_x^2+n_y^2+n_z^2$ with $1 \leq n_x \leq n_{\mathrm{max}}$, $1
\leq n_y \leq \sqrt{n_{\mathrm{max}}^2 - n_x^2}$, and $1 \leq n_z \leq
\sqrt{n_{\mathrm{max}}^2 - n_x^2- n_y^2}$; this list had all the
energies $\epsilon \leq n_{\mathrm{max}}^2$ (we chose
$n_{\mathrm{max}}=32$). This is a very degenerate system---of the
15,954 energy levels only 818 are distinct and the average
degeneracy is 19.50. As seen in Fig.~\ref{fig:3DHistobig}, for $\epsilon \sim
1000$ we have $d_\epsilon \sim 60$.  Specifically, the energies from 933--954,
along with their degeneracies, are given by:
$\{\epsilon,\, g(\epsilon)\}$ = \{933,\,24\}, \{934,\,39\},
\{936,\,24\}, \{937,\,27\}, \{938,\,24\}, \{\,\}, \{940,\,6\},
\{941,\,66\}, \{942,\,18\}, \{944,\,9\}, \{945,\,48\}, \{946, \,24\},
\{947,\,15\}, \{948,\,18\}, \{949,\,12\}, \{950,\,63\}, \{952, \,12\},
\{953,\,45\}, \{954,\,42\}.

\emph{Step 3}: Plot ${\mathcal N}(\epsilon)$. It is instructive to
construct the set of points for the plot explicitly using the list
of energies. There are only 818 steps, one for each value of
$\epsilon$. The height of the step at some energy $\epsilon$ is
$d_\epsilon$, the degeneracy, so this staircase has steps that get
higher with energy (e.g., the step at $\epsilon=950$ has a height of 63).

\begin{figure}[h!]
\includegraphics[width=8.5cm]{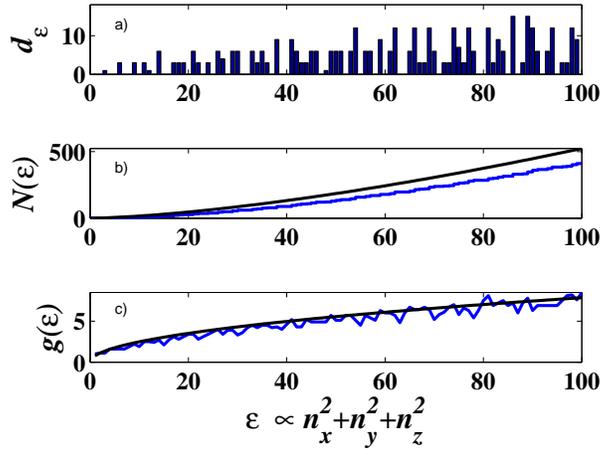}
\caption{\label{fig:3DHistosmall} Top: Degeneracy $d_\epsilon$ of the
allowed energies $\epsilon$ (in units of $\epsilon_{0}$) of a single
particle in a 3-D rigid cubic box, up to $n_x=n_y=n_z=10$. Middle: The
corresponding ${\mathcal N}(\epsilon)$ is compared to the
theoretical result (smooth line) ${\mathcal N}(\epsilon)=
(\pi/6)\epsilon^{3/2}$. Bottom: A plot of $g(\epsilon)$
obtained using the derivative scheme in Eq.~(\ref{eqn:deriv}) with a
window width of 10. Theory (smooth line) gives
$g(\epsilon)=(\pi/4)\epsilon^{1/2}$.}
\end{figure}

\begin{figure}[h!]
\includegraphics[width=8.5cm]{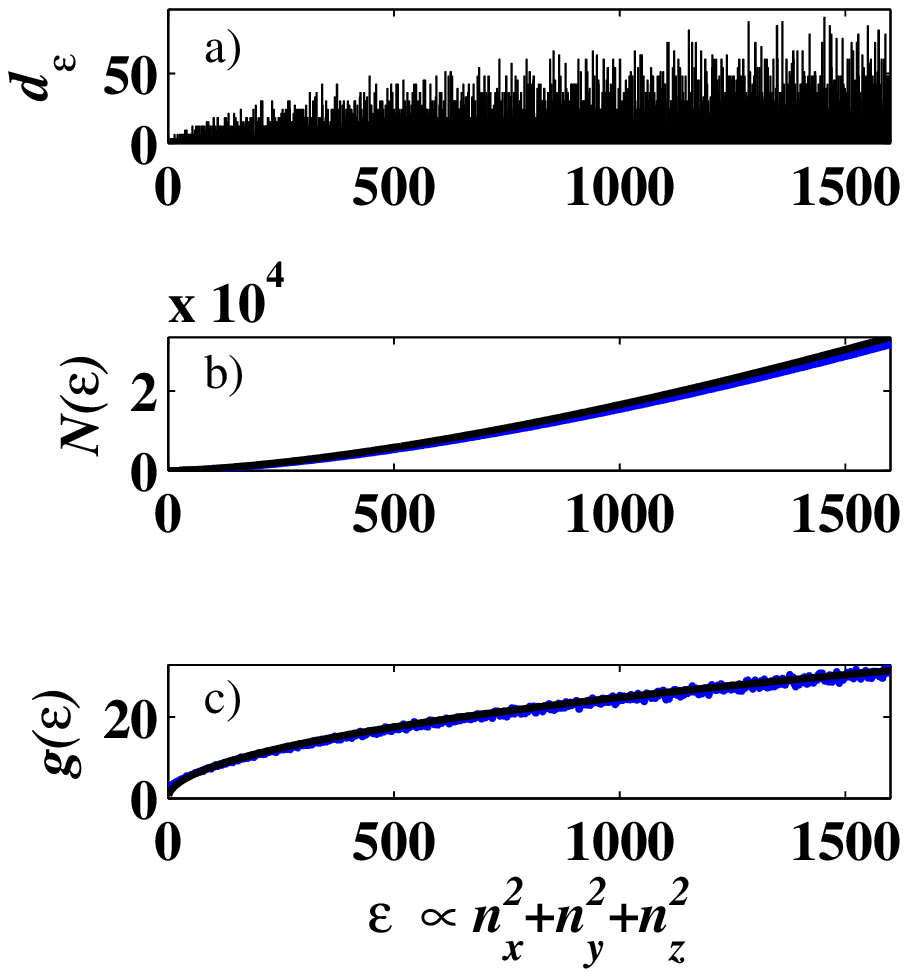}
\caption{\label{fig:3DHistobig} Top: Degeneracy $d_\epsilon$ of the
allowed energies $\epsilon$ (in units of $\epsilon_{0}$) of a single
particle in a 3-D cubical box, up to $n_x=n_y=n_z=40$. Middle: The
corresponding ${\mathcal N}(\epsilon)$ (lower curve) is compared to the
theoretical result (smooth line, upper) ${\mathcal N}(\epsilon)=
(\pi/6)\epsilon^{3/2}$. Bottom: A plot of $g(\epsilon)$
obtained using the derivative scheme in Eq.~(\ref{eqn:deriv}) with a
window width of 50. Theory (smooth line) gives
$g(\epsilon)=(\pi/4)\epsilon^{1/2}$.}
\end{figure}

\emph{Step 4}:  While calculating the DOS, the window size needs to
be adjusted to give a reasonable-looking graph. If the window is too
small, the graph will look noisy; if it is too big, the graph may
appear too coarse. We use trial and error to select an appropriate window size. We
have included code in the online supplement\cite{EPAPS} in which the reader can select the
maximum energy and window size and generate plots like
Figs.~\ref{fig:3DHistosmall} and \ref{fig:3DHistobig}.

Figures~\ref{fig:3DHistosmall} and \ref{fig:3DHistobig} display
$d(\epsilon)$, ${\mathcal N}(\epsilon)$, and $g(\epsilon)$ for this
system. Again, it is evident that the agreement between the analytic result and the
explicit counting of energies improves as the energy interval
increases.

\subsection{\label{sec:rectangle}Rectangular box}

The rigid rectangular box is a straightforward adjustment to
the cubical case and gives us a look at a non-degenerate spectrum. In
order to remove the degeneracies from the single-particle
spectrum that exist in the cubical case, we will make the
sides of the box have incommensurate lengths.  The choice $L_x\neq L_y
\neq L_z$ with $L_x=1$, $L_y=2/e$, and
$L_z=e/2$ gives us a box of unit volume and a
non-degenerate spectrum.

\emph{Step 1}: The energies are $\epsilon=(\pi^2\hbar^2)[(n_x/L_x)^2+(n_y/L_y)^2+(n_z/L_z)^2]/(2M)$.
Again, the obvious choice for our energy scale is to set
$\pi^2 \hbar^2/(2M) = 1$.

\emph{Step 2}:  This time we use a different method
than in the cubical case. The sphere in $n$-space corresponds to an
ellipse in phase space. This is a nice illustration of the
difference between these abstract spaces.  The method used
was an exhaustive calculation of all the energies
corresponding to the points $1 \leq n_x \leq n_{\mathrm{max}}$, $1
\leq n_y \leq n_{\mathrm{max}}$, and $1 \leq n_z \leq n_{\mathrm{max}}$, after which we
selected $\epsilon \leq (2 n_{\mathrm{max}} /e)^2$. The
fact that $n_y$ and $n_z$ are multiplied by a factor of 0.74
and 1.36 respectively ensured that the sphere radius of
$\sqrt{2/e}\,n_{\mathrm{max}}$ was enclosed in the ellipse in
phase space, and we had a complete set of energies. Although
the method is inefficient, it has the advantage of
transparency.

All the energies in this system are unique. In order to have
degeneracies we would need two sets of integers, $\{n_x,n_y,n_z\}$
and $\{m_x,m_y,m_z\}$, such that ${n_x}^2+e^2
n_y^2/4 + 4 n_z^2/e^2 = {m_x}^2+e^2
{m_y}^2/4 + 4{m_z}^2/e^2$, which is impossible as $e$ is
transcendental. The lowest 62,440 energies of the system were
calculated.

\emph{Step 3}: The cumulative state number ${\mathcal
N}(\epsilon)$ consists of steps of unit height, though the
widths of the steps (level spacings) vary.  For our
rectangular box there are 41,435 steps, one for each energy.
A plot of $d_\epsilon$ vs\ $\epsilon$ is a series of horizontal spikes
of height~1.

\begin{figure}[h!]
\includegraphics[width=8.5cm]{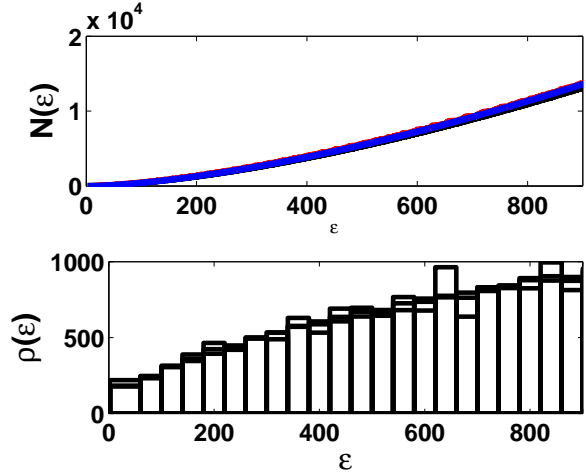}
\caption{\label{fig:rhoNsingleparticle} The upper panel shows the
cumulative state number for the spectra for a single particle in
cubical, rectangular, and spherical boxes. The spectra have been
normalized so that $\hbar^2\pi^2/(2M)=1$ and the volume of
each box is 1. The three curves are essentially indistinguishable.
The lower panel is a histogram of the energies for each box. The
bins are 40 units wide.}
\end{figure}

\emph{Step 4}: In Fig.~\ref{fig:rhoNsingleparticle} we show
$g(\epsilon)$ and ${\mathcal N}(\epsilon)$ for this system.
[xx Fig. 7 has a label of $\rho$ which has not been defined. xx]


[xx LEFT OFF HERE xx]

\subsection{\label{sec:sphere}
Spherical box}

The energies for a rigid sphere of radius $R$ are
$\epsilon_{ln}=\hbar^2k_{ln}^2/(2 M R^2)$, where $k_{ln}$ is
the $n$th zero of $j_l(r)$, the $l$th
spherical Bessel function.\cite{LiboffBook}  Each $\epsilon_{ln}$
has a degeneracy of $2l+1$.  It is easy to confuse $j_l(r)$ with
$J_l(r)$, the ordinary Bessel function, also known as the Bessel
function of the first kind; the two are related by \cite{LiboffBook}
$j_l(r)=\sqrt{\pi/(2r)}J_{l+1/2}(r)$. Note that the
$n$th zero of $J_l(r)$ is written $K_{ln}$.
{\sc Mathematica} has a useful add-on called
\verb=NumericalMath`BesselZeros'= that lists the zeros of the
various Bessel functions. In  Sec.~\ref{sec:problems} we define the
problem of the rigid cylindrical box, where the energies are
proportional to $K_{ln}^2$, so to avoid confusion we provide the
following check: the first three zeros of $J_0(r)$ are
$\{2.4048,5.5201,8.6537\}$, and the first three zeros of $j_0(r)$ are
$\{3.9374, 7.8748, 11.8122\}$. Actually, the distinction between
$K_{ln}$ and $k_{ln}$ is not important as far as $g(\epsilon)$ goes
for the sphere and the cylinder, as the two sets of zeros are
interspersed.\cite{Liboff}

\emph{Step 1}: The role of $\pi$ in the energies of the rectangular
wells is taken over by the $k_{ln}$. In the rectangular box  the
zeros of the sine function are multiples of $\pi$. This allowed us
to factor it out and choose the energy scale to be in units of
$\hbar^2 \pi^2/(2M)$ for the rectangular boxes. Now we have $\epsilon
= \hbar^2 (k_{ln}/R)^2/(2M)$. To compare
with the square wells we use an energy scale with $\hbar^2
\pi^2/(2M)=1$; in these units, $\epsilon_{ln}=[k_{ln}/(\pi
R)]^2$. Also, if the volume of the box is to be unity, then
$R=[3/(4\pi)]^{1/3}$, so finally we have
$\epsilon_{ln}=k_{ln}^2/(3 \pi^2/4)^{2/3}$.
This is the spectrum we will compare to those of the rectangular and
cubical boxes.

\emph{Step 2}: We calculated all energies with $k_{ln} < 95$. To ensure a
complete set, we used the brute force method and calculated all the
triplets $\{k_{ln},\,l,\,n\}$ with $0\leq l \leq 95$ and $1\leq n \leq
95$, sorted these 9,120 numbers, and selected the first 95. Making
the triplets  as opposed to just the list of $k_{ln}$ made it easy to
accommodate the $2l+1$ degeneracy when making the final spectrum.

\emph{Steps 3 \& 4}: See Fig.~\ref{fig:rhoNsingleparticle}.


\subsection{\label{sec:weyl} Weyl's theorem}

\begin{figure}[h!]
\includegraphics[width=8.5cm]{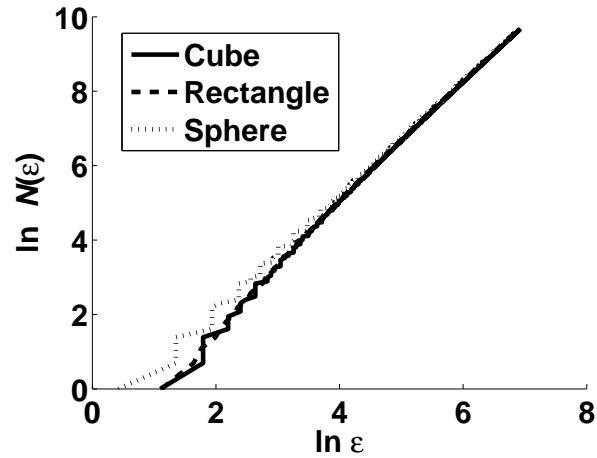}
\caption{\label{fig:logNsingleparticle}A log-log plot of the
cumulative state number for the single-particle spectra of cubical,
rectangular, and spherical boxes. The energies are in units of
$\hbar^2\pi^2/(2M)$ and the volume of each box is~1.}
\end{figure}

It is clear from Sec.~\ref{sec:box} and
Fig.~\ref{fig:rhoNsingleparticle} that the general trend for the three
boxes we examined is ${\mathcal N}(\epsilon)=\alpha\,
\epsilon^{\beta}$. The energy exponent does not vary much with the
shape of the box. A plot of $\ln{\mathcal N}(\epsilon)$ vs\
$\ln\epsilon$ illustrates this nicely, as shown in
Fig.~\ref{fig:logNsingleparticle}. The deviation from the straight
line occurs only at low energies, where the shape of the box
matters. Using {\sc Mathematica} we performed linear fits to the log-log plots
and obtained the results shown in Table~\ref{tab:enE}.

If the box is sufficiently large, a particle in a rigid box
should not be aware of the particular shape of the box,
so long as $\lambda^D \ll V$, where
$V$ is the volume of the box. Here, $\lambda = 2
\pi/|\vec{k}|$ is the deBroglie wavelength ($|\vec{p}|=\hbar
|\vec{k}|$) and $D$ is the dimension of the space.
A slow particle will know about the edge, as it has a
long wavelength, and a fast particle  or high-energy state
will not be sensitive to the shape because its wavelength is small. This is the content of Weyl's theorem,\cite{Weyl} which can be paraphrased as ``high energy eigenvalues of the wave equation are insensitive to the shape of the boundary.''  A nice account of the origin of the theorem is given by Kac,
\cite{Kac} and an explicit proof at the level of this
paper for the case of the cylinder and sphere is given by
Lambert.\cite{Lambert}


\section{\label{sec:npart}$N$ particles: statistical mechanics}

Statistical mechanics can be introduced by analyzing a model system of $N$ non-interacting particles on a single-particle spectrum, often chosen as a set of equally-spaced levels. \cite{Serway} This is a valuable pedagogical tool for exploring the statistical distributions for classical particles, identical bosons, and identical fermions.  This type of model is rich with subtleties: the particles not only don't see each other beyond obeying the Pauli exclusion principle if they are identical fermions, but their presence has no influence on the energies of the levels they fill.\cite{Noninteracting} Real particles, on the other hand, have interactions, so how does this model give useful results? One can think of the effect of all the other particles as being a smooth mean field that can be modeled as a potential for that single particle. Another subtlety is the robustness of the spectra of these $N$-body systems: they are very insensitive to the details of the single-particle spectrum upon which they are built, depending only on its DOS.

We made a simple model of the the $N$-particle system to serve as a
starting point for students to explore what happens to the DOS as $N$ increases, and to illustrate that the DOS of many-body systems like this are insensitive to the details of the single-particle spectra they are based on. In this section $E$ is used for the total energy of the system, reserving
$\epsilon$ for the single-particle energy. Subscripts will be used
on $g$ and ${\mathcal N}$ to indicate the number of
particles in the system, so $g_7(E)$ is the DOS for a system
of seven particles. We compared $N$-particle systems built on four distinct single-particle spectra: the square, rectangular, and spherical 3-D boxes, and a spectrum with a specified DOS but random levels.  A simple analytical model for the DOS is presented and compared with the numerical results.  The focus is on the relationship between the
original single-particle energies that the $N$ particles populate
and the resulting $g_N(E)$. The model demonstrates that $g_N(E)$
depends on the \textit{density} of the single-particle spectrum, as
opposed to the details of the energies themselves. This is why we
can replace the exact single-particle energies (zeros of Bessel functions, etc.)
with any set of numbers as long as they have the same range and
density.

Using a simple algorithm, the spectrum of $N$ identical bosons on a set of single-particle energies $\{\epsilon_i \}$ was calculated. The range of energies was $0 \rightarrow E_{\mathrm{max}}$, with $E_{\mathrm{max}}$ limited only by
computational power and user patience.  The choice of bosons avoided the subtleties of Fermi statistics. The algorithm is best illustrated with an example.  Let's look at identical bosons on the single-particle energies of the rigid cube in three dimensions.  First we make a list of the single-particle-state energies: $\{\epsilon_i\}=\{3, 6, 6, 6, 9, 9, 9, 11, 11, 11, 12, 14, 14,\dots\}$.  We choose $E_{\mathrm{max}}=53$ so the list includes the first 140 states. This means that we are guaranteed to have a complete list of energies up to $E=53$.  For our purposes an $N$-particle state is a list of occupied single-particle states, so $(4,5,5,27)$ is a four-particle state with a particle in state 4, two particles in state 5, and a particle in state~27. Now list the indices of all possible one-particle states: $\{(i)\}=\{(1),(2),(3),\dots,(140)\}$; the energy of state $(i)$ is $\epsilon_i$. Next list the index pairs of all possible two-particle states: $\{(i,j)\}=\{(1,1),(1,2),(1,3),\dots,(140,140)\}$, sort them and drop duplicates; the energy of the state $(i,j)$ is $\epsilon_i+\epsilon_j$. Drop all states with $E>53$ as you are only guaranteed a complete set of energies for $E\leq 53$ . In general, list all possible $N$-particle states, sort them, drop duplicates, and sum the corresponding single-particle energies. Once you have these spectra you can use the procedure in Sec.~\ref{sec:procedure} to get $g_N(E)$ and ${\mathcal N}_N(E)$.

Systems based on a random single-particle spectrum were included for comparison with those based on the rigid boxes because this removed the role of geometry. The distribution of the random numbers was such that the DOS was $g(\epsilon)=(\pi/4)\epsilon^{1/2}$; this was accomplished by taking a set
$\{x_i\}$ of 500 random numbers uniformly distributed on the
interval $[0,1]$, sorting them, raising them to the power of
$2/3$, and multiplying them by $[(6/\pi)500]^{2/3}$. The logic behind this is worth mentioning. Given a set of numbers $\{x\}$ with a
distribution $f(x)$, we can make a function $y(x)$ such
that the  set $\{ y(x) \}$ has a distribution $\alpha \,
y^\beta $ by noting that $f(x)\,dx =\alpha\,y^\beta\,dy$.
In our case  $f(x)=1$, and we have $dy/dx=(1/\alpha)y^{-\beta}$ so that $y(x) = [(\beta +1)/\alpha]^{1/(\beta +1)}x^{1/(\beta +1)}$.

\begin{figure}[h!]
\includegraphics[width=8.5cm]{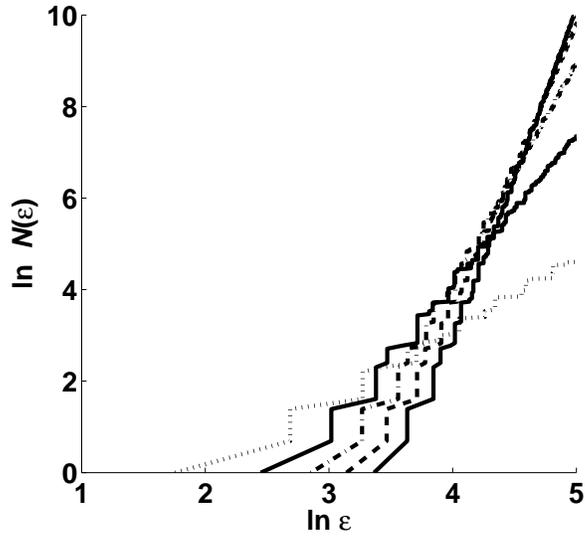}
\caption{\label{fig:lnNlnE}A log-log plot of the cumulative state
number for $N$-particle systems ($N=1,2,\dots, 5$) on the single-particle spectrum of the sphere. The steeper lines correspond to
higher values of $N$, and different line types have been used for
clarity. The slopes follow $N(\beta + 1)$. }
\end{figure}

\begin{figure}[h!]
\includegraphics[width=8.5cm]{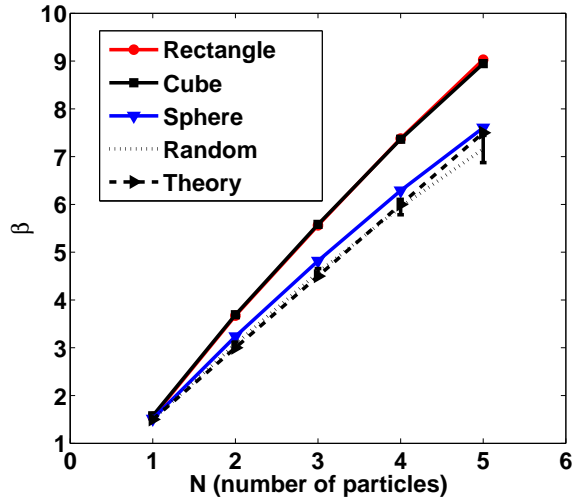}
\caption{\label{fig:beta} The exponent of $E$ in ${\mathcal
N}(E)=\alpha E^\beta$ for systems of various particle number. Each
system is made by putting non-interacting particles on the
corresponding single-particle spectrum described in the text. The dashed line is a theoretical value derived using the values
$a=0.4$ and $b=1/2$ in Eq.~(\ref{eqn:npart}). These results were
obtained using a linear fit to the $\log {\mathcal N(E)}$ data.
(The line is a guide for the eye.) The error bars represent the
standard deviation of the average values from the 100 random
spectra.}
\end{figure}

\begin{figure}[h!]
\includegraphics[width=8.5cm]{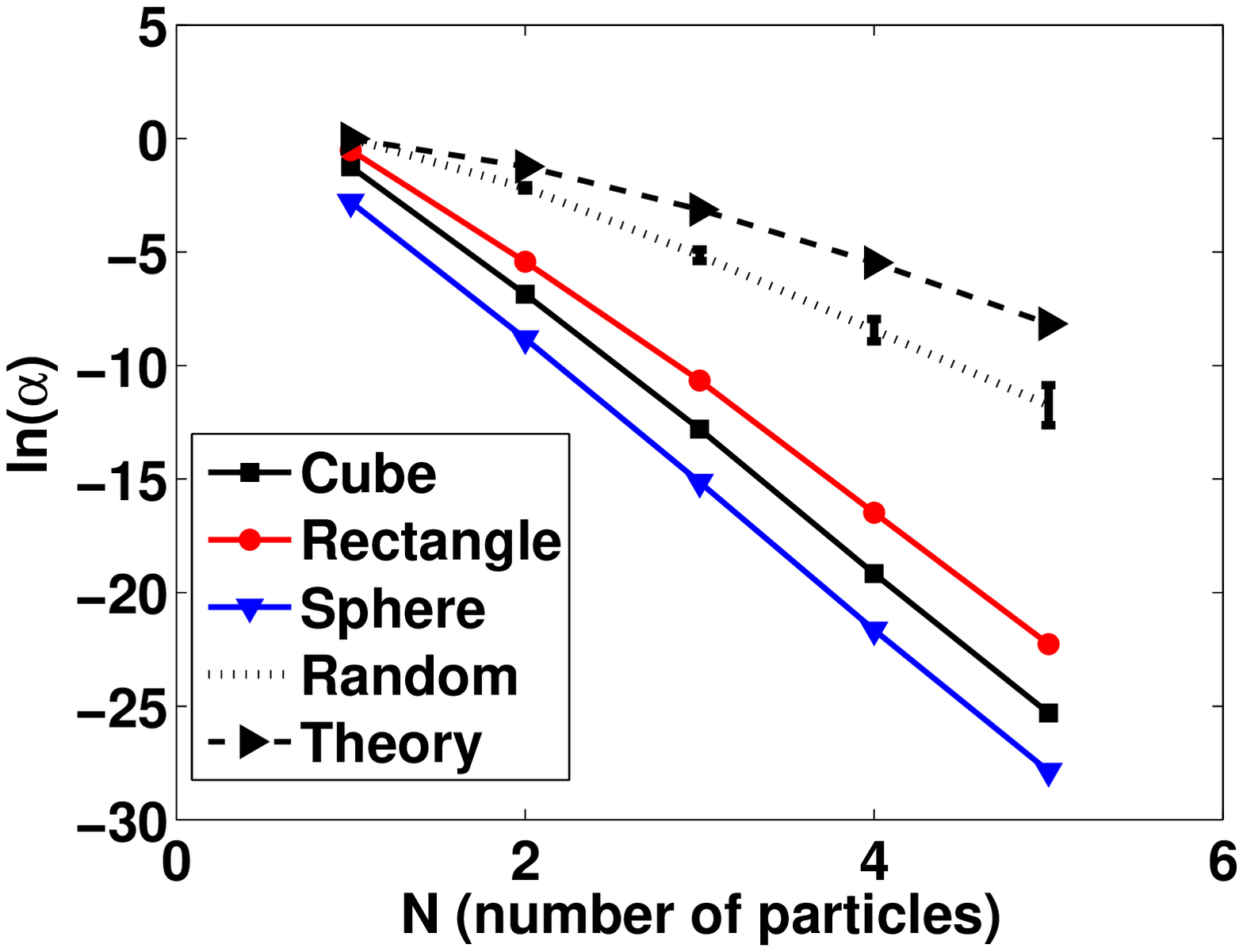}
\caption{\label{fig:alpha} A graph of $\ln\alpha$ vs\ $N$ for the
same systems as in Fig.~\ref{fig:beta}. Again, these results were obtained
using a linear fit to the $\ln {\mathcal N(E)}$ data. The error bars
represent the standard deviations of the average values from the 100
random spectra.}
\end{figure}

All of these systems had a cumulative state number of the form ${\mathcal
N}_N(E)=\alpha E^\beta$, with both $\beta$ and  $\ln \alpha$ being
linear in $N$. In Fig.~\ref{fig:lnNlnE} we show ${\mathcal N}_N(E)$
vs\ $E$ [xx Fig. 9 actually plots vs $\epsilon$ instead of $E$.  Please correct wording as necessary. xx]
for $N$ particles on the single-particle energies of a
spherical well, while Figs.~\ref{fig:beta} and~\ref{fig:alpha} show
graphs of $\beta$ and $\ln \alpha$ as functions of $N$ for the
various systems we examined. The values of $\alpha$ and $\beta$ for the random spectra were based on 100 separate single-particle spectra.

To compare with the theory, we used
$g_1(\epsilon)=0.40 \epsilon^{1/2}$.  [xx I don't understand why this sentence is here. Should it be part of the next paragraph?  xx]

Given a single-particle spectrum with a one-particle DOS
$g_1(E)$, we can obtain a naive expression for the $N$-body
density of states $g_N(E)$ iteratively. The density of
$N$-body states with energy $E$ is a product of the density
of ($N-1$)-body states with energy $E'$ and one-body states with
energy $E-E'$, or
\begin{equation}
\label{eqn:rhon} g_N(E)=\int_0^Eg_{N-1}(E')g_1(E-E')\,dE'.
\end{equation}
This expression  double-counts some states. Suppose we add to the
three-particle state $\{n_1,n_2,n_3\}=\{3, 17, 17\}$ a
fourth particle in state 24; then $E=\epsilon_3+2
\epsilon_{17}+\epsilon_{24}$, and there is a contribution to the
integral  in Eq.~(\ref{eqn:rhon}) from
$g_1(\epsilon_{24})g_3(\epsilon_3+2 \epsilon_{17})$.  However, we
can get to the same state by adding a particle in state 17 to the
three-particle state $\{n_1,n_2,n_3\}=\{3, 17, 24\}$, and this gives a
contribution to the integral in Eq.~(\ref{eqn:rhon}) from
$g_1(\epsilon_{17})g_3(\epsilon_3+ \epsilon_{17}+\epsilon_{24})$.
This is the origin of the double counting and its effect is to
increase the number of states by a number close to $N$ at high
energies, where the occupancies of the single-particle states are
low. We did not fix this problem. When iterating Eq.~(\ref{eqn:rhon})
we need the following result for integer $n$  and $m$:
\begin{equation}
\label{eqn:conv} \int_0^E(E')^n\,
(E-E')^m\,dE'\,=\frac{n!m!}{(n+m+1)!}E^{n+m+1}.
\end{equation}
Table~\ref{tab:rhon} gives several examples of the DOS for $N$
bosons based on single-particle spectra with level density $g_1(E)=
E^b$, for $b = -1/2,\,0,\,1/2,\,1$, corresponding to
a box in 1, 2, 3, and 4 dimensions.


Immediately we see some trends. The DOS for one boson in a
$p$-dimensional rigid box, Eq.~(\ref{eqn:gepd}), is the same as
that of $p$ bosons in a one-dimensional rigid box [xx You see this by comparing what to what? xx]. This
result is a by-product of separation of variables in the Cartesian case.
However, if we put the particles into a spherical box, we
also have separation of variables, but they [xx What is ``they'' referring to? xx] are not
equivalent. The $a$ in $g_1(E)=a E^b$ changes, but the powers
of $g_N(E)$ do not [xx It is not clear where this result comes from. xx]. In general, upon iterating
Eq.~(\ref{eqn:rhon}), for $N$ bosons on a single-particle
spectrum with $g_1(E)=a E^b$, we get
\begin{eqnarray}
\label{eqn:npart}
   g_N(E)&=& \frac{b!^N}{(Nb+N-1)!}\, a^N E^{Nb+N-1}, \\
\label{eqn:npart2}{\mathcal N}_N(E)&=& \frac{b!^N}{(Nb+N)!}\, a^N
E^{Nb+N}=\alpha E^{\beta}.
\end{eqnarray}
This simple treatment exhibits the properties
of the systems we have examined [xx Exhibits what properties?  This sentence seems incomplete. xx]. The results in Table~\ref{tab:rhon}
and Figs.~\ref{fig:lnNlnE}--\ref{fig:alpha} illustrate this [xx Illustrate what perfectly? xx]
perfectly. In Figs.~\ref{fig:beta} and \ref{fig:alpha} we used the
values $a=0.4$ and $b=1/2$ in Eq.~(\ref{eqn:npart2}) for comparison.
In Eq.~(\ref{eqn:npart2}) we see that $\ln(\alpha)$ is the sum of $N
\ln(a)$ and a term that depends on $b$, so  it is not very sensitive
to the value of $a$. The theoretical value of $\alpha$ is higher
than for any of the systems examined. This is partially due to the
double counting discussed above, which makes ${\mathcal N}_N(E)$
larger, but that may not be the whole story. In the case of
$\ln(\alpha)$ and $\beta$ the random system was closest to the
theory, but we will not speculate as to why.

[xx I found this last section, and in particular the last few paragraphs to be very confusing overall.  I really lost sight of what you are trying to show.  Please consider revising some of this section to try to clarify things as much as possible. xx]

\section{\label{sec:conclusion}
Conclusions}

We have provided a graphical approach to introduce students to the
density of states (DOS) for simple quantum systems. In
Sec.~\ref{sec:intro} we introduced the concepts of degeneracy,
cumulative state number ${\mathcal N}(\epsilon)$, and DOS
$g(\epsilon)$ for these systems.  In Sec.~\ref{sec:procedure} we
gave a procedure for visualizing the DOS and used it to get the DOS
for a particle in rigid boxes of various shapes. We found that
the shape of the box had little effect on the DOS, a
demonstration of Weyl's theorem. We then built $N$-particle systems
on the single-particle spectra of Sec.~\ref{sec:procedure} and found that the
DOS was insensitive to the specific energies of the single-particle
spectra, the single-particle DOS ($g_1(\epsilon)$) being the
important factor. Finally, we compared our numerical results for ${\mathcal
N}_N(E)$ with an expression based on iterating the single-particle
density of states and saw how the DOS for $N$-particle systems was
simply related to $g_1(\epsilon)$.

\section{\label{sec:problems}
Additional Problems}

The following problems provide additional practice with the procedure used in Section~III.
The programs available in Ref.~\onlinecite{EPAPS} may be a helpful starting point.

\begin{enumerate}
\item Examine changes to the density of
states if we continuously change one of the dimensions of the
rigid rectangular box.
We expect the density
of states to exhibit different behavior as we change
continuously from a ``2-D'' system ($L_z \ll L_{x},L_{y} $),
to a ``3-D'' system ($L_z\approx L_{x},L_{y}$) to a ``1-D''
system ($L_z \gg L_{x},L_{y} $). Let $L_x = L_y = L_{0}= 1$, and then let
$L_z$ vary over the range $ 0.01 L_{0}\leq L_{z} \leq 100 L_{0}$.

\item The energy levels for a particle in a rigid cylindrical box of height
$H$ and radius $R$ are given by\cite{LiboffBook}
$\epsilon_{qln}=\hbar^2(q^2\pi^2/H^2 +
K_{ln}^2/R^2)/(2 M)$, where the $K_{ln}$ are
zeroes of the (regular) Bessel functions.
Vary the height-to-radius ratio for the cylindrical box. As
$H/R$ goes from near 0 to 1, you should expect a
transition from ``2-D'' behavior to ``3-D'' behavior. What does this
look like in terms of $g(\epsilon)$?

\item Redo the treatment of Sec.~\ref{sec:box} for a relativistic particle
in the 2-D rigid square box.  The important difference is
that relativistically $\epsilon=pc=\hbar k c$, where $k$ is the wave number
and $c$ is the speed of light.
\end{enumerate}

\begin{acknowledgments}
We thank T. Bensky for helpful discussions.
For their hospitality during sabbatical leaves M.~J.~M.\ thanks the
faculty of the School of Physics at the Dublin Institute of
Technology and D.~M.\ is
grateful to the physics department at Temple University.
The reviewers provided detailed comments that significantly
clarified our presentation.
\end{acknowledgments}

\newpage

\section*{Tables}
\begin{table}[ht!]
\caption{\label{tab:dims}
The DOS for a particle in a rigid box of $D$ dimensions. The
energy $\epsilon$ is in units of the lowest single particle
energy $\epsilon_0$.}
\begin{center}
  \begin{tabular}{@{} ccccccc @{}}
  \toprule
  $D$ & ${\mathcal N}(\epsilon) $& $g(\epsilon)=d{\mathcal N}/d\epsilon$ \\
\hline
  1 & $\epsilon^{1/2} $ & $\frac{1}{2} \,  \epsilon^{-1/2} $\\
  2 & $\frac{\pi}{4}\epsilon$ & $\frac{\pi}{4}$\\
  3 & $\frac{\pi}{6} \, \epsilon^{3/2}$ & $\frac{\pi}{4} \, \epsilon^{1/2}$\\
  4 & $\frac{\pi^2}{32} \, \epsilon^2$ & $\frac{\pi^2}{16} \, \epsilon$\\
  5 & $\frac{\pi^2}{60} \, \epsilon^{5/2}$ & $\frac{\pi^2}{24} \, \epsilon^{3/2}$\\
  6 & $\frac{\pi^3}{384} \, \epsilon^{3}$ & $\frac{\pi^3}{128} \, \epsilon^{2}$\\
  \botrule
  \end{tabular}
\end{center}
\end{table}

\begin{table}[ht!]
  \caption{\label{tab:enE}
Parameters for ${\mathcal N}(\epsilon)=\alpha \epsilon^\beta$ for
the three different 3-D ``boxes.'' The theory gives
$\alpha=\pi/6\approx 0.524$ and $\beta=3/2$; see Table~\ref{tab:dims}. }
\begin{center}
  \begin{tabular}{@{} lcc @{}}
  \toprule
  $ $ & $\alpha$ & $\beta$  \\ \hline
  Cube (theory) & $\pi/6\approx 0.524$ & $\quad 3/2=1.5$ \\
  Cube (numerical) &   0.29 & 1.58  \\
  Rectangular box &  0.32   &  1.56   \\
  Spherical box &   0.46 &  1.51\\
  \botrule
  \end{tabular}
  \end{center}
\end{table}

\begin{table*}[ht!]
  \caption{\label{tab:rhon}
The DOS for $N$-particle systems built on single-particle spectra
with a DOS $g_1(E)=a\,E^b$. We set $a=1$ for simplicity but $g_N(E)$
includes a factor of $a^N$.}
\begin{center}
  \begin{tabular}{@{} ccccc @{}}
  \toprule
  $ $ & $b=-1/2$ & $b=0$ & $b=1/2$ & $b=1$ \\ \hline
  $g_1(E)$ &  $E^{-1/2}$ & $1$  & $E^{1/2}$& $E$\\
  $g_2(E)$ &  $\pi$  &  $E$   &  $\frac{\pi}{8}E^2$& $\frac{1}{3!} E^3$\\
  $g_3(E)$ &  $2 \pi E^{1/2}$   &  $\frac{1}{2!} E^2$ &  $\frac{2\pi}{105}E^{7/2}$& $\frac{1}{5!} E^5$\\
  $g_4(E)$ &  $\pi^2 E$   &  $\frac{1}{3!} E^3$ &  $\frac{\pi^2}{1920}E^5$& $\frac{1}{7!} E^7$\\
  $g_5(E)$ &  $\frac{4}{3}\pi^2 E^{3/2}$    &  $\frac{1}{4!} E^4$ &  $\frac{4 \pi^2}{135135}E^{13/2}$& $\frac{1}{9!} E^9$\\
  $g_6(E)$ &  $\frac{1}{2}\pi^3 E^2$    &  $\frac{1}{5!} E^5$ &  $\frac{\pi^3}{2580480}E^8$& $\frac{1}{11!} E^{11}$\\
  $g_7(E)$ &  $\frac{8}{15}\pi^3 E^{5/2}$    &  $\frac{1}{6!} E^6$ &  $\frac{8 \pi^3}{654729075}E^{19/2}$& $\frac{1}{13!} E^{13}$\\
  $g_8(E)$ &  $\frac{1}{6} \pi^4 E^3$   &  $\frac{1}{7!} E^7$ &  $\frac{\pi^4}{10218700800}E^{11}$& $\frac{1}{15!} E^{15}$\\
  $\vdots $ &    &   &  & \\
  $g_N(E)$ &  $\frac{\pi^{N/2}}{(N/2-1)!} E^{N/2-1}$  &  $\frac{1}{(N-1)!} E^{N-1}$ &  $\frac{\pi^{N/2}}{2^N (3N/2-1)!} E^{3N/2-1}$& $\frac{1}{(2N-1)!} E^{2N-1}$\\
  \botrule
  \end{tabular}
\end{center}
\end{table*}

\clearpage

\section*{Figure captions}

\begin{thebibliography}{99}
\bibitem{Noninteracting}Anecdotally, one of
the authors (D.~M.) has been surprised by the persistent
student misconception that ``fermions don't interact."  This false
notion can be traced to the commonly used  ``system of
noninteracting fermions" that introduces statistical
mechanics in the intermediate course. The equidistant single-particle spectrum is a common starting point when looking at
the spectra of many-body systems. It is confusing that after
all the work invested in solving the single-particle
Schr\"odinger equation, we just throw away our hard won
zeros-of-Bessel-functions-based energies, and replace them
with the integers.
\end{thebibliography}
\end{document}